\documentclass{article}
\usepackage[T1]{fontenc}
\usepackage{iclr2027_conference,times}

\usepackage{amsmath,amsfonts,bm}

\def\eqref#1{equation~\ref{#1}}

\def\1{\bm{1}}

\DeclareMathAlphabet{\mathsfit}{\encodingdefault}{\sfdefault}{m}{sl}
\SetMathAlphabet{\mathsfit}{bold}{\encodingdefault}{\sfdefault}{bx}{n}

\usepackage{amsmath}
\usepackage{graphicx}
\usepackage{amssymb}
\usepackage{amsthm}
\usepackage{algorithm}
\usepackage{algpseudocode}
\usepackage{booktabs}
\usepackage{array}
\usepackage{xcolor}
\usepackage{hyperref}
\hypersetup{hidelinks}
\usepackage{url}
\usepackage{tikz}
\theoremstyle{definition}
\newtheorem{definition}{Definition}
\newtheorem*{example}{Example}
\theoremstyle{plain}

\newtheorem{proposition}{Proposition}

\newcommand{\dmax}{d_{\max}}

\title{New Snake-in-the-Box Records via Snakepit Surgery and Learned Construction}
\author{\parbox[t]{\dimexpr\textwidth-2\tabcolsep}{\raggedright
  Paul~Orland\textsuperscript{1} \quad
  Lucas~Fagan\textsuperscript{1} \quad
  Michele~Tarquini\textsuperscript{1} \quad
  Davide~Passaro\textsuperscript{1} \quad
  Maksymilian~Manko\textsuperscript{2} \quad
  Elli~Heyes\textsuperscript{3} \quad
  Angus~Gruen \quad
  Giorgi~Butbaia\textsuperscript{1} \quad
  Justin~Tan\textsuperscript{4} \quad
  Sergei~Gukov\textsuperscript{1}\\[4pt]
  {\normalfont\small
  \textsuperscript{1}California~Institute~of~Technology \quad
  \textsuperscript{2}Universit\"at~Z\"urich \quad
  \textsuperscript{3}Imperial~College~London \quad
  \textsuperscript{4}University~of~Cambridge}}
}
\iclrfinalcopy
\begin{document}
\maketitle
\lhead{}
\renewcommand{\headrulewidth}{0pt}
\raggedbottom

\begin{abstract}
The snake-in-the-box problem asks for a longest induced path in the hypercube graph $Q_n$. We find a length-$191$ snake in dimension $n=9$---the lowest dimension where the maximum is unknown---improving the previous record of $190$ that had stood for 14 years. We also establish new lower bounds in dimensions $10$--$13$. To find these records, we introduce snakepits, collections of disjoint snakes, to expand the search space and open new routes between snakes. This motivates our new Snakepit-in-the-Box benchmark, which seeks maximal edge counts when allowing multiple components. Finally, we introduce Beam Anchor, a search-supervised learned constructor algorithm that finds $100$ inequivalent length-$190$ snakes in dimension $9$.
\end{abstract}

\section{Introduction}
The hypercube graph $Q_n$ has binary strings of length $n$ as vertices and edges between strings that differ in exactly one coordinate. A \emph{snake} is an induced (chordless) path: consecutive vertices are adjacent and distinct nonconsecutive vertices are not. The snake-in-the-box problem, introduced by \citet{kautz}, asks for the maximum snake length, $a(n)$. 
The problem came from coding theory as error-detecting Gray codes, and it is an ideal test of combinatorial search:
snakes are easy to construct and validate, but long snakes are exceedingly rare and the optimal length is known only through dimension $8$ \citep{ostergard}.

There are two complementary ways to search for long snakes.

\emph{Construction} grows a snake one vertex at a time; each choice restricts later continuations from visiting adjacent vertices. This is a long-horizon decision problem in which the cost of poor decisions may not be realized until hundreds of moves later, so immediate reward is uninformative and credit assignment is difficult.

\begin{figure}[h]
    \centering
    \includegraphics[width=0.9\linewidth]{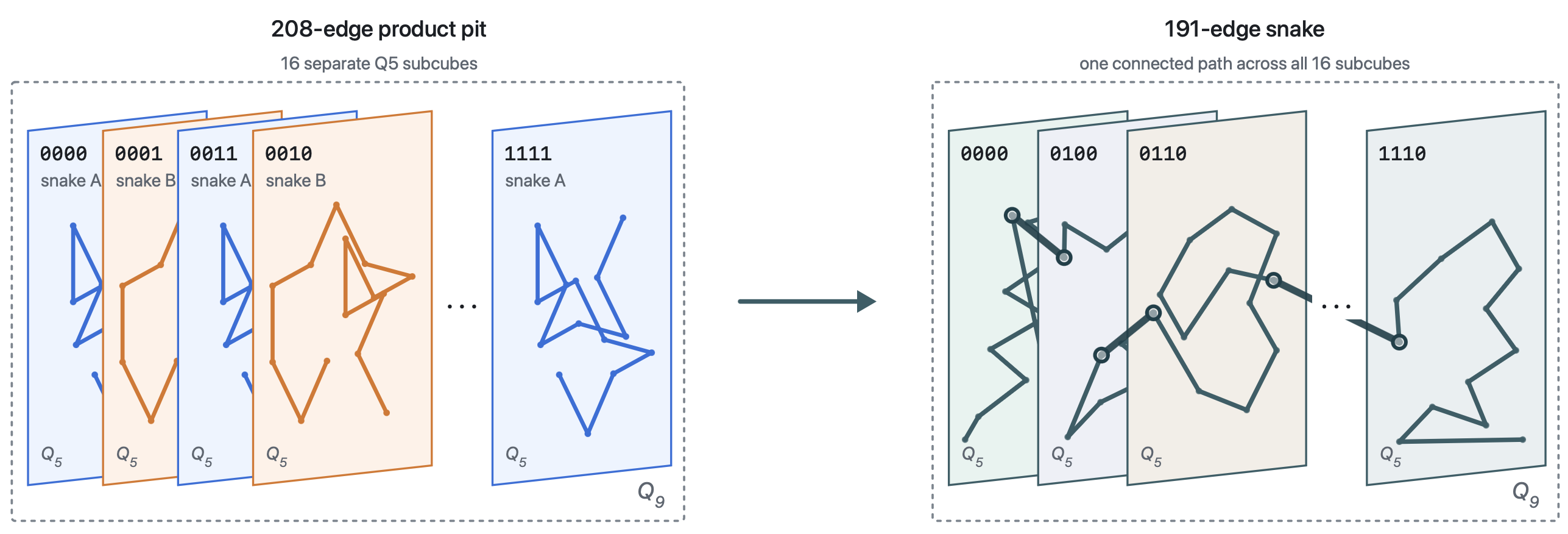}
    \caption{From a snakepit to a snake. Two vertex-disjoint length-$13$ snakes in $Q_5$ can be combined to form a $16$-snake snakepit with $208$ edges in $Q_9$. Cube-surgery annealing can rearrange such a starting state into a length-$191$ snake.}
    \label{fig:pit-to-snake}
\end{figure}

\emph{Modification}, on the other hand, makes local changes to an existing snake. However, interior routes are highly constrained for long snakes, making them difficult to change, since most vertices are invalidated by being adjacent to a different part of the snake. We propose best-in-class techniques for both approaches.

For modification, we gain flexibility by expanding the state space to \emph{snakepits}, collections of disjoint snakes, which allows for easier routes between snakes and therefore makes rewards less sparse. Our algorithm uses a basic move called \emph{cube surgery}: clear a subcube and rebuild its contents subject to the surrounding snakes. Allowing snakepits makes cube surgery a more flexible operation: a refill can change the overall connectivity structure of the snakepit. We discovered the new record snakes in dimension 9 using simulated annealing with cube surgery with a penalty for larger number of components.

For construction, we introduce Beam Anchor, which combines beam search with a learned policy to construct a single path by comparing continuations from a common prefix. From a partial snake, it gives each possible next move its own share of search and trains the policy to favor moves whose continuations reach longer snakes. 

In sum, our main contributions are:
\begin{itemize}
\item \textbf{New records, led by $191$ in $Q_9$.} We find a length-191 snake in dimension 9, breaking the record of 190 that had stood for 14 years, and improve snake bounds in dimensions 10--13 alongside new coil bounds. Our bank of 131 inequivalent length-191 snakes supports an empirical study of family structure and search coverage.
\item \textbf{Snakepit search with cube surgery.} We relax the search state from a single snake to a snakepit and rebuild small subcubes in one annealing proposal. At equal time this beats add/remove annealing by about $40$ edges in $Q_9$ and rediscovers the record from an empty cube.
\item \textbf{Learned construction with Beam Anchor.} We apply reinforcement learning to snake construction and develop Beam Anchor, which finds $100$ inequivalent length-$190$ snakes in $Q_9$, greatly expanding the previous known snake examples at the longstanding record length.
\item \textbf{Snakepit-in-the-Box.}  We introduce a new benchmark that naturally follows from our use of snakepits as intermediate search states: maximizing the edge count of a snakepit in a given dimension. We provide lower bounds, exact values through $Q_7$, and a density limit.

\end{itemize}

Sections~\ref{sec:modification} and \ref{sec:construction} develop modification and construction; we then discuss the records, landscape and coils, and close with the snakepit benchmark.
\begin{table}[t]
\centering
\caption{Lower bounds established by this work relative to the records preceding our announcement. The final column counts inequivalent snakes found at the record length, up to the symmetries of $Q_n$ and reversal.}
\vspace{0.2cm}

\label{tab:records}
\begin{tabular}{c c c c c}
\toprule
dimension $n$ & our bound $a(n) \geq$ & previous record & $\Delta$ & \# classes at record\\
\midrule
$9$  & $\mathbf{191}$  & $190$ \citep{wynn}           & $\mathbf{+1}$   & $131^{*}$ \\
$10$ & $\mathbf{379}$  & $376$ \citep{ace2025}        & $\mathbf{+3}$   & $\geq 12$ \\
$11$ & $\mathbf{746}$$^{\dagger}$  & $737$ \citep{ace2026}        & $\mathbf{+9}$   & $\geq 72$ \\
$12$ & $\mathbf{1476}$$^{\dagger}$ & $1465$ \citep{ace2026}       & $\mathbf{+11}$  & $\geq 34{,}688$ \\
$13$ & $\mathbf{2924}$$^{\dagger}$ & $2900$ \citep{echolsfurther} & $\mathbf{+24}$  & $\geq 30$ \\
\bottomrule
\end{tabular}
\par\smallskip
{\footnotesize $^{*}$Discovered bank, not a census; see Section~\ref{sec:landscape}.\par
$^{\dagger}$Subsequent reported bounds: $750$ \citep{echols2026}, $1480$ and $2934$ \citep{itty2026}.\par}
\end{table}

\section{Related Work}
\label{sec:related}
Exhaustive search established the snake optima through $Q_8$ \citep{kochut,ostergard}. Higher-dimensional records have come from population-based search \citep{potter,casella}, permutation-based constructions \citep{wynn}, Monte-Carlo search \citep{kinny,meyerson2015}, nested rollout policy adaptation \citep{dang}, and beam search and local rerouting \citep{ace2025,ace2026,echolswindowed,echolsfurther}. Our higher-dimensional bounds are historical improvements; Table~\ref{tab:records} marks subsequent advances.

Cube surgery connects penalty-based annealing \citep{kirkpatrick} with ruin-and-recreate and large-neighborhood search \citep{schrimpf,shaw}. Collections of separated snakes also occur in coil enumeration \citep{ostergardcoil}; here they are intermediate states that permit repeated interior rearrangement. Beam Anchor is policy-only expert iteration: beam search supplies the teacher and policy distillation guides later search \citep{anthony}. It shares the search--learning loop of AlphaZero \citep{alphazero}, but uses no learned value head and supervises only the anchor where alternatives were compared.

\section{Snakes and Snakepits}
\label{sec:defs}
Vertices of $Q_n$ are binary strings, with coordinates indexed from $0$ to $n-1$. An edge flips one coordinate. A $d$-dimensional subcube leaves $d$ coordinates free and fixes the rest.

\begin{definition}[Snakes and coils]
A snake in $Q_n$ is an induced path $(v_0,\ldots,v_L)$ of distinct vertices. Consecutive vertices are adjacent and nonconsecutive vertices are not. Its length is $L$, and it has $L+1$ vertices. We write $a(n)$ for the maximum snake length. A \emph{coil} is an induced cycle; its length equals both its edge count and its number of distinct vertices.
\end{definition}
The \emph{transition sequence} records the coordinate flipped at each step.
We consider snakes up to translations, coordinate permutations and reversal: to canonicalize a transition sequence, relabel coordinates by order of first appearance, do the same to the reversed sequence (and cyclic permutations for coils), and take the lexicographically smallest result.
A class count always refers to this equivalence unless explicitly stated otherwise.

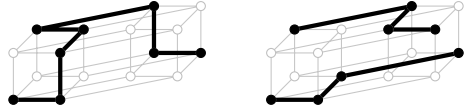
\begin{figure}[h]
\centering
\begin{tikzpicture}[
    scale=0.62,
    vertex/.style={circle,draw=gray!50,fill=white,inner sep=1.2pt}
]

\begin{scope}[xshift=0cm]

  \coordinate (d0) at (0,0);      \coordinate (d1) at (1,0);
  \coordinate (d2) at (0,1);      \coordinate (d3) at (1,1);
  \coordinate (d4) at (0.5,0.5);  \coordinate (d5) at (1.5,0.5);
  \coordinate (d6) at (0.5,1.5);  \coordinate (d7) at (1.5,1.5);

  \coordinate (d8)  at (2.5,0.5); \coordinate (d9)  at (3.5,0.5);
  \coordinate (d10) at (2.5,1.5); \coordinate (d11) at (3.5,1.5);
  \coordinate (d12) at (3.0,1.0); \coordinate (d13) at (4.0,1.0);
  \coordinate (d14) at (3.0,2.0); \coordinate (d15) at (4.0,2.0);

  \draw[gray!45]
    (d0)--(d1) (d0)--(d2) (d0)--(d4)
    (d1)--(d3) (d1)--(d5)
    (d2)--(d3) (d2)--(d6)
    (d3)--(d7)
    (d4)--(d5) (d4)--(d6)
    (d5)--(d7) (d6)--(d7)
    (d8)--(d9) (d8)--(d10) (d8)--(d12)
    (d9)--(d11) (d9)--(d13)
    (d10)--(d11) (d10)--(d14)
    (d11)--(d15)
    (d12)--(d13) (d12)--(d14)
    (d13)--(d15) (d14)--(d15)
    (d0)--(d8) (d1)--(d9) (d2)--(d10) (d3)--(d11)
    (d4)--(d12) (d5)--(d13) (d6)--(d14) (d7)--(d15);

  \draw[line width=1.6pt]
    (d0)--(d1)--(d3)--(d7)--(d6)--(d14)--(d12)--(d13);

  \foreach \i in {0,...,15}
    \node[vertex] at (d\i) {};
  \foreach \i in {0,1,3,7,6,14,12,13} \node[vertex,draw=black,fill=black] at (d\i) {};

\end{scope}

\begin{scope}[xshift=5.5cm]

  \coordinate (e0) at (0,0);      \coordinate (e1) at (1,0);
  \coordinate (e2) at (0,1);      \coordinate (e3) at (1,1);
  \coordinate (e4) at (0.5,0.5);  \coordinate (e5) at (1.5,0.5);
  \coordinate (e6) at (0.5,1.5);  \coordinate (e7) at (1.5,1.5);

  \coordinate (e8)  at (2.5,0.5); \coordinate (e9)  at (3.5,0.5);
  \coordinate (e10) at (2.5,1.5); \coordinate (e11) at (3.5,1.5);
  \coordinate (e12) at (3.0,1.0); \coordinate (e13) at (4.0,1.0);
  \coordinate (e14) at (3.0,2.0); \coordinate (e15) at (4.0,2.0);

  \draw[gray!45]
    (e0)--(e1) (e0)--(e2) (e0)--(e4)
    (e1)--(e3) (e1)--(e5)
    (e2)--(e3) (e2)--(e6)
    (e3)--(e7)
    (e4)--(e5) (e4)--(e6)
    (e5)--(e7) (e6)--(e7)
    (e8)--(e9) (e8)--(e10) (e8)--(e12)
    (e9)--(e11) (e9)--(e13)
    (e10)--(e11) (e10)--(e14)
    (e11)--(e15)
    (e12)--(e13) (e12)--(e14)
    (e13)--(e15) (e14)--(e15)
    (e0)--(e8) (e1)--(e9) (e2)--(e10) (e3)--(e11)
    (e4)--(e12) (e5)--(e13) (e6)--(e14) (e7)--(e15);

  \draw[line width=1.6pt]
    (e0)--(e1)
    (e1)--(e5)
    (e5)--(e13)
    (e6)--(e14)
    (e14)--(e10)
    (e10)--(e11);

  \foreach \i in {0,...,15}
    \node[vertex] at (e\i) {};
  \foreach \i in {0,1,5,6,10,11,13,14} \node[vertex,draw=black,fill=black] at (e\i) {};

\end{scope}

\end{tikzpicture}
\caption{A snake with $7$ edges (left) and a snakepit with $6$ edges in two components (right), both in $Q_4$.}
\label{fig:q4-comparison}
\end{figure}

\begin{definition}[Snakepit]
A snakepit $F$ is an induced linear forest in $Q_n$. Equivalently, it is a collection of (possibly length 0) snakes with no cube edge between vertices in distinct components.  Write $v(F)$, $c(F)$ and $e(F)$ for its numbers of vertices, components and edges, respectively. Since each component is a path, we have that $e(F)=v(F)-c(F)$.
\end{definition}
A snake, then, is a snakepit with one component. Importantly, the intersection of a snake(pit) with a subcube is also a snakepit: deleting vertices can split paths but cannot make nonconsecutive vertices adjacent. A sample snake and snakepit in $Q_4$ are given in Figure~\ref{fig:q4-comparison}.

The \emph{Snakepit-in-the-Box} problem asks for the maximum edge count of a snakepit in $Q_n$. Its finer breakdown by component count is
\[
 L(n,k)=\max\{e(F):F\text{ is a snakepit in }Q_n,\ c(F)=k\},
\]
for feasible $k\geq1$. In particular, $L(n,1)=a(n)$. Section~\ref{sec:snakepit} develops the table of values and bounds.

\begin{definition}[Product snakepit]
\label{def:product}
 Let $A$ and $B$ be snakepits in $Q_m$ with no common vertices and let $n>m$. In $Q_n=Q_{n-m}\times Q_m$, place a copy of $A$ over each even-parity vertex of $Q_{n-m}$ and a copy of $B$ over each odd-parity vertex.
\end{definition}
Clearly the result is a snakepit satisfying
\[
 e(F)=2^{n-m-1}\bigl(e(A)+e(B)\bigr),\qquad
 c(F)=2^{n-m-1}\bigl(c(A)+c(B)\bigr).
\]
\begin{example}[Product snakepits in $Q_9$]
Two disjoint length-$13$ snakes in $Q_5$ give a pit with $16$ components, $208$ edges and $224$ vertices in $Q_9$, represented schematically on the left of Figure~\ref{fig:pit-to-snake}. A disjoint length-$26$ and length-$24$ pair in $Q_6$ gives $8$ components, $200$ edges and $208$ vertices in $Q_9$. These pits contain more edges than a length-$191$ snake, but spread them across multiple components.
\end{example}

\section{Finding Record Snakes: Modification}
\label{sec:modification}
Cube-surgery annealing found our length-$191$ snake in $Q_9$. Local rerouting has contributed to previous records \citep{ace2025,echolswindowed}, but long snakes leave little room for interior edits: a replacement must avoid vertices adjacent to the rest of the snake while reconnecting the pieces it separates. We relax the connectivity requirement and search over snakepits, allowing an edit to leave several components that can later be joined into a longer snake.

\paragraph{Cube surgery.}
\label{sec:surgery}
Cube surgery rebuilds a small subcube while holding every exterior vertex fixed. Each proposal has four steps:
\begin{enumerate}
\item \textbf{Clear a region.} Save the current state, draw $d$ uniformly from $\{2,3,4\}$, choose a uniform $d$-subcube $U$, and remove every occupied vertex inside it. These $4$--$16$-vertex regions permit coordinated changes at low refill cost.
\item \textbf{Prefer connections.} Visit the unoccupied vertices of $U$ in random order. Immediately add a vertex if it has at least one occupied neighbor and the addition is legal. This pass extends or joins existing fragments without starting a new component.
\item \textbf{Allow new fragments.} Make a fresh random pass, now allowing any legal addition, including a vertex with no occupied neighbors. Such a vertex starts a new component that later additions can grow. Both passes test against the current state, including vertices just added. Repeat the two passes until a full round adds nothing.
\item \textbf{Accept or restore.} Compare the energy of the rebuilt state with the saved original and apply the annealing acceptance rule below once to the whole edit. On rejection, restore the original state. Individual deletions and insertions receive no separate acceptance decisions.
\end{enumerate}
A legal addition starts a component, extends a path at an endpoint, or joins two paths at their endpoints, so every refill is again a snakepit. Appendix~\ref{detail:sec:surgery} gives pseudocode.

\paragraph{Annealing and consolidation.}
To balance occupied vertices against fragmentation, the annealer uses a component penalty. For a snakepit $F$ and penalty $\beta>0$, it minimizes
\begin{equation}
 E(F)=\beta|c(F)-1|-v(F).
 \label{eq:energy}
\end{equation}
For a single snake of length $L$, this is $-(L+1)$; the empty state has energy $\beta$. A proposed change $\Delta E$ is accepted with probability $\min\{1,\exp(-\Delta E/T)\}$. The temperature repeatedly decreases from $T_0$ toward $T_{\min}$, then reheats. Hot phases permit fragmentation and temporary loss of vertices; cooling favors consolidation. Because the proposal is asymmetric, this is an optimization heuristic, not a claim of Boltzmann sampling.

\paragraph{Other search moves.}
Alongside surgery, the annealer proposes single-vertex additions (with the same legality rule) and removals, which can shorten or split a path.

Endpoint moves replace a longer segment in one proposal: a \emph{wiggle} retracts $j\in\{1,\ldots,6\}$ vertices from a path endpoint and regrows up to $j+2$ steps through legal neighbors in random order, and deep regrowth does the same with $8$--$48$ vertices retracted and up to $12$ extra steps. Acceptance is evaluated only after regrowth. 

\paragraph{From product snakepit to record snake.}
The first successful chains started from the product snakepits of Section~\ref{sec:defs}. Although these pits contain more edges than a $191$-snake, their fragmentation is penalized by Eq.~\ref{eq:energy}: surgery rearranges the fragments while annealing favors their consolidation. The product start is a convenience rather than a requirement, since chains started from the empty snakepit also reach $191$, as shown next.

\paragraph{The value of surgery.}
We compared two fixed recipes in $Q_9$, each started from the empty snakepit and run for one hour on one CPU core, over ten paired seeds: single-vertex addition and removal alone, and the same moves with surgery proposed $35\%$ of the time. Both recipes allow fragmented states, so the comparison isolates the move rather than the relaxation. After one hour, the add/remove recipe had reached a median best length of $149$ (range $147$--$155$) and the surgery recipe $188$ (range $186$--$191$); every paired difference favored surgery, by $35$ to $41$ edges, even though the cheaper recipe made about $2.4\times$ more proposals in the same time. One surgery run reached a validated $191$ after $158$ seconds. Appendix~\ref{app:surgery-comparison} gives the protocol, all twenty runs and the progress at earlier checkpoints. The same annealer also refines the lifted snakes described next.

\section{Finding Record Snakes: Construction}
\label{sec:construction}
Our successful ordinary beam-search pipeline uses a long lower-dimensional snake as a prefix to build long snakes. Reusing such paths is an established way to make construction tractable. Beam Anchor addresses the harder task of constructing a competitive snake from the origin, without a supplied primer. We first explain the primer strategy, then how Beam Anchor discovers long snakes from scratch. 

\subsection{Beam search with a primer}
\label{sec:beam}
Beam search grows valid partial snakes one edge at a time. At each depth, it generates legal extensions, ranks the candidates, and retains a bounded number for the next step. Our lifter ranks candidates by length plus the number of reachable available vertices and Gumbel noise; this favors continuations with room to grow while diversifying the search.

Starting from a strong primer reduces the part of the path that search must discover. We embed a long snake from $Q_{n-1}$ in one half of $Q_n$ as a fixed prefix, then grow a continuation into the other half. This generally produces a snake that crosses the new coordinate exactly once, although unrestricted continuation may cross back. Reusing lower-dimensional snakes is a common way to make record searches tractable \citep{kinny,wynn}, but it can miss optima: the known length-$98$ snake in $Q_8$ uses no coordinate exactly once, and extending an optimal length-$50$ $Q_7$ primer cannot reach $98$ \citep{kinny}.

We use ordinary beam search for lifting because it is substantially faster than Beam Anchor, making it practical to extend many record primers. We try both primer orientations and selected tail truncations. Each lift remains a single path, and its output seeds cube-surgery annealing (Section~\ref{sec:modification}). Thus the pipeline alternates beam-search construction in a larger cube with modification of the resulting snake. The observed raw lifts $379\to743$, $746\to1470$, and $1476\to2916$ show how improvements propagate between dimensions. The full scoring rule, widths, and deduplication details appear in Appendix~\ref{detail:sec:construction}.

\subsection{Beam Anchor}
\label{sec:beam-anchor}
Beam Anchor is a learned constructor: it builds a snake from the origin one edge at a time, with no primer, guided by a policy network. It follows the expert iteration scheme \citep{anthony}: a search procedure acts as an expert that improves on the current policy, the policy is trained to imitate the expert, and the improved policy in turn guides the next round of search. In Beam Anchor the expert is a set of policy-guided beam searches, one for each legal move at the current partial snake, and the policy learns which move led to the longest snake. There is no value network.

In $Q_9$, Beam Anchor finds $100$ inequivalent snakes of length $190$. Before this project, Wynn's construction was the only published length-$190$ class known to us \citep{wynn}, so these snakes substantially enlarge the published landscape at the previous record length; all $100$ also occur in the bank of $1{,}042$ length-$190$ classes produced by our annealer (Section~\ref{sec:landscape}). To our knowledge, it is also the first learned constructor to reach length $190$ in $Q_9$, and it does so with no prior knowledge of the problem: it starts from the origin with a randomly initialized policy, and neither the search nor the training data draw on primers, lower-dimensional snakes or previously known snakes. Everything the policy knows it has learned from its own searches.

\paragraph{Why compare openings.}
Immediate reward is uninformative for snake construction: every legal move adds exactly one edge, but the result of a move and the vertices it excludes only becomes consequential many steps later. The key question at a partial snake is therefore global: which of the legal next moves admits the longest completion? Beam Anchor answers this by searching at every step of a construction and distilling the answers into a policy that guides later searches.

\paragraph{Anchor search.}
Let $p$ be a partial snake, the \emph{anchor}, and let $A(p)$ be its legal next moves, $m=|A(p)|$. A fixed budget of $B$ beams is divided evenly among the legal moves, and we run one beam search per move $a\in A(p)$, starting from the child $p\mathbin\Vert a$, the partial snake $p$ extended by the move $a$, with width $\lfloor B/m\rfloor$ or $\lceil B/m\rceil$. The searches never compete for beams, so every move is explored with the same effort regardless of how the current policy ranks it. Within a search, candidates are ranked by cumulative policy log-probability perturbed by Gumbel noise, and the search runs until every beam has completed, that is, reached a state with no legal move. Its outcome is the longest snake it completed,
\[
 \widehat L(p,a)=\max\{\ell(s): s \text{ a completed snake found by the search through } a\},
\]
where $\ell$ is edge length. This provides a lower bound on the best completion through $a$.

\paragraph{Walks.}
A \emph{walk} builds a snake from the origin one move at a time, running an anchor search at each step and sampling the next move from its outcomes.
At the current anchor $p$ it (i) runs the anchor search, (ii) records the outcomes in a persistent table,
\begin{equation}
 \mathcal M[p,a]\leftarrow\max\{\mathcal M[p,a],\widehat L(p,a)\}\qquad\text{for all } a\in A(p),
 \label{eq:anchor-memory}
\end{equation}
and (iii) samples its next move from $\mu(a\mid p)\propto\exp\bigl(\widehat L(p,a)/\tau_{\mathrm{walk}}\bigr)$ and moves the anchor to $p\mathbin\Vert a$. The walk ends when the anchor has no legal move.

A walk may complete a longer snake than the anchor search found at some prefix along its path, and one could credit that length to the move the walk took there. We do not: that move is followed by further anchor searches while the other moves at the prefix are not, so it has effectively received more search budget, and crediting it would bias the targets toward the moves the walk happened to choose rather than the moves that are better.

\paragraph{Distillation.}
A cycle runs a batch of walks in parallel and then updates the policy on rows sampled from $\mathcal M$. For a row $p$, the target distribution and loss are
\begin{equation}
 q_\tau(a\mid p)=\frac{\exp(\mathcal M[p,a]/\tau)}{\sum_{u\in A(p)}\exp(\mathcal M[p,u]/\tau)},
 \qquad
 \mathcal L(\theta)=-\mathbb E_p\sum_{a\in A(p)}q_\tau(a\mid p)\log\pi_\theta(a\mid p),
 \label{eq:distillation-loss}
\end{equation}
where $\pi_\theta(\cdot\mid p)$ is the policy's distribution over the legal moves $A(p)$. Rows are drawn by sampling a depth uniformly and then a prefix at that depth; otherwise the numerous deep prefixes would crowd out the few early ones, where mistakes are most costly. The temperature $\tau$ is higher at shallow depths, where a bounded search separates openings least reliably. The updated policy guides the anchor searches of the next cycle (Algorithm~\ref{detail:alg:beam-anchor}, Appendix~\ref{app:beam-anchor}).

The policy is the only learned component. A value network would have to predict, from a prefix, the length of its best completion, which may lie more than a hundred moves ahead; and to be useful at an anchor it would have to do so to within a few edges, since that is all that separates the best completions of rival moves. The policy target is easier: it only asks how the moves at one prefix rank against each other.

\paragraph{Policy network.}
The policy is a graph neural network on $Q_n$ itself, with one node per vertex and messages along cube edges. Each vertex $v$ starts from a vector $x_v\in\mathbb R^8$ of features of the partial snake $p$: whether $v$ is on the snake, still available, the start, the head, or a legal next vertex; its normalized position along the snake; and the fractions of the vertices at Hamming distance one, and at distance two, from $v$ that lie on the snake. Six residual GIN layers of width $64$ turn these into embeddings $h_v\in\mathbb R^{64}$. Let $v$ be the head of $p$, and let $v\oplus e_a$ be the vertex reached by the move $a$, where $e_a$ is the unit vector of coordinate $a$. The logit of $a$ is
\[
 g_\theta(a\mid p)=\mathrm{MLP}\bigl([h_{v},\,h_{v\oplus e_a}]\bigr),\qquad
 \pi_\theta(a\mid p)=\frac{\exp g_\theta(a\mid p)}{\sum_{u\in A(p)}\exp g_\theta(u\mid p)},
\]
with one MLP shared by all moves (Appendix~\ref{app:beam-anchor}). No feature or weight refers to a coordinate label, so the network is equivariant under every automorphism of $Q_n$ by construction. 
\paragraph{Experiments.}
\emph{Training.} We trained two different policies from two random seeds for $12$ hours on an eight-device MI250 node. Beam Anchor first reached length $190$ in $Q_9$ after about $8$ and $8.5$ hours, respectively, and continued finding new $190$-snakes as training continued. Without learning, Beam Anchor with a uniform policy plateaus at $178$ (Appendix~\ref{app:anchor-protocol}).

\emph{Search with the trained policy.} The trained policies remain useful without further training. With their parameters frozen, we ran Beam Anchor walks from the origin for $12$ hours on one eight-device node, with four random seeds (two for each policy). In total, across training and these frozen-policy runs, Beam Anchor found $100$ inequivalent length-$190$ snakes.

\emph{Comparison with PPO.} 
To assess the importance of the search, we compare Beam Anchor to a PPO \citep{ppo} baseline with the same graph neural network (Appendix~\ref{app:anchor-protocol}). Across all PPO experiments, the best snake found had length 184.

\section{Records and Search Landscape}
\label{sec:records}
\paragraph{Record provenance.}
\label{sec:pipeline}
Table~\ref{tab:records} reports the improvements, led by $191$ in $Q_9$, which came from cube-surgery annealing alone. Lifting and annealing independently reached $379$ in $Q_{10}$. In $Q_{11}$--$Q_{13}$, lifts reached $743$, $1470$ and $2916$; annealing took them to $746$ and $1476$, and in $Q_{13}$ a genetic search with mixed-integer linear programming (MILP) window refill, which fixes everything outside a window of the path and solves exactly for the longest induced path through it, supplied the $2921$ seed for the final annealing step to $2924$. In Appendix~\ref{app:data} we give a $Q_9$ and $Q_{10}$ snake at record length; all other data and class banks are available at \url{https://github.com/Math-AI-Caltech/Snake-in-the-Box}.

\subsection{What the \texorpdfstring{$Q_9$}{Q9} Bank Reveals About Search}
\label{sec:landscape}
To describe the $131$ inequivalent $191$-snakes, the \emph{dimension-crossing vector} (DCV) counts uses of each coordinate, sorted in decreasing order. Its entries sum to the snake's length; its largest entry, $\dmax$, measures how many times the most-used dimension appears.

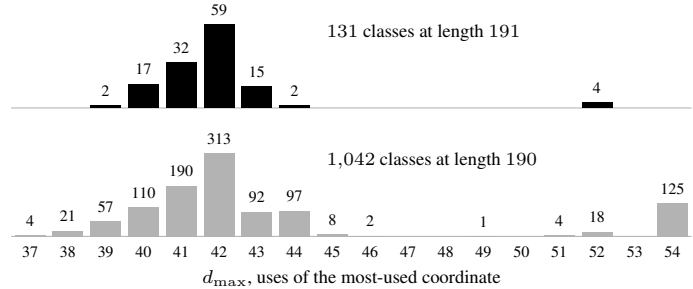
\begin{figure}[h]
\centering
\begin{tikzpicture}[x=0.5cm,y=1cm]
\begin{scope}[yshift=1.7cm]
\draw[gray!60] (36.5,0) -- (54.5,0);
\foreach \x/\c in {39/2,40/17,41/32,42/59,43/15,44/2,52/4}{
  \fill[black] (\x-0.4,0) rectangle (\x+0.4,{\c/59*1.1});
  \node[font=\tiny,above] at (\x,{\c/59*1.1}) {\c};
}
\node[font=\scriptsize,anchor=west] at (44.6,1.0) {$131$ classes at length $191$};
\end{scope}
\draw[gray!60] (36.5,0) -- (54.5,0);
\foreach \x/\c in {37/4,38/21,39/57,40/110,41/190,42/313,43/92,44/97,45/8,46/2,49/1,51/4,52/18,54/125}{
  \fill[gray!60] (\x-0.4,0) rectangle (\x+0.4,{\c/313*1.1});
  \node[font=\tiny,above] at (\x,{\c/313*1.1}) {\c};
}
\node[font=\scriptsize,anchor=west] at (44.6,1.0) {$1{,}042$ classes at length $190$};
\foreach \x in {37,...,54}{\node[font=\tiny,below] at (\x,0) {\x};}
\node[font=\scriptsize] at (45.5,-0.55) {$\dmax$, uses of the most-used coordinate};
\end{tikzpicture}
\caption{Distribution of $\dmax$ over the known $Q_9$ classes at length $191$ (top) and $190$ (bottom). At length $191$ the continent ($\dmax\leq44$) and the island ($\dmax=52$) are separated by an empty gap; the length-$190$ stratum shows the same two modes.}
\label{fig:dmax}
\end{figure}

The $191$-snakes separate into a \emph{continent} of $127$ classes with $\dmax\in\{39,\ldots,44\}$ and an \emph{island} of four with $\dmax=52$; no class has $\dmax$ between $45$ and $51$ (Figure~\ref{fig:dmax}). 
Both families are basins. From a neutral start, such as a product or the empty snakepit, annealing lands in either family depending on the seed; once there, it readily finds the rest of that family but never reaches the other (Appendix~\ref{app:landscape}). This separates two evaluation questions: how thoroughly a method explores a known family, and whether it reaches a different one. Beam Anchor illustrates the distinction: all $100$ of its length-$190$ classes have $\dmax\leq46$, in the continent's mode, and none lies in the island's (Figure~\ref{fig:ba-dmax}, Appendix~\ref{app:beam-anchor}). None of the $131$ known $191$s admits a one-edge extension, so a $192$-snake requires a class outside this bank.

\subsection{Coils}
\label{sec:coils}
The same searches, adapted, set new coil records (Table~\ref{detail:tab:coils}). A coil is an induced cycle (Section~\ref{sec:defs}); a \emph{symmetric} coil $(v_0,\ldots,v_{2h-1})$ also satisfies $v_{i+h}=v_i\oplus t$ for a fixed $t$. The records came from three routes.
\begin{table}[t]
\centering
\small
\caption{Coil and symmetric-coil bounds established by this work. Class counts are up to the symmetries of $Q_n$, reversal and rotation.
The separate symmetric-coil rows refer to translation symmetry.}
\label{detail:tab:coils}
\begin{tabular}{@{}c c c c c @{}}
\toprule
$n$ & our bound & previous & $\Delta$ & \# classes at record \\
\midrule
\multicolumn{5}{@{}l}{\emph{Coils}} \\
$9$  & $\mathbf{192}$  & $188$ \citep{wynn}           & $+4$  & $1$ \\
$10$ & $\mathbf{374}$  & $370$ \citep{echolswindowed} & $+4$  & $12$  \\
$11$ & $\mathbf{734}$$^{\dagger}$  & $728$ \citep{ace2026}        & $+6$  & $9$  \\
$12$ & $\mathbf{1466}$$^{\dagger}$ & $1442$ \citep{ace2026}       & $+24$ & $5$  \\
$13$ & $\mathbf{2916}^{\dagger\ddagger}$ & $2840$ \citep{echolsfurther} & $+76$ & $13$ \\
\midrule
\multicolumn{5}{@{}l}{\emph{Symmetric coils}}  \\
$10$ & $\mathbf{370}$ & $362$ \citep{meyerson2015}  & $+8$ & $2$  \\
$11$ & $\mathbf{734}$ & $718$ \citep{echolsquotient} & $+16$ & $1$ \\
\bottomrule
\end{tabular}
\par\smallskip
{\footnotesize $^{\dagger}$Subsequent reported bounds: coils $740$, $1468$, $2930$ in dimensions $11$--$13$ \citep{echols2026,minortriad}.\par
$^{\ddagger}$Found by MILP window refill from a length-$2932$ snake of Echols, separately from the harvest-and-close campaign.\par}
\end{table}
\paragraph{Harvest and close ($n=9$--$12$).} A snake whose endpoints are at distance two closes into a coil two longer if a common neighbor of the endpoints is unused and adjacent to no other path vertex. In $Q_9$, $96$ of the $1{,}042$ length-$190$ classes close, all to the same $192$-coil. In $Q_{10}$--$Q_{12}$ no record-length snake closes, so we ran the annealer warmer with a lowered save threshold and tested the $14$--$24$ million shorter snakes it harvested per dimension, giving $374$, $734$ and $1466$.

\paragraph{MILP window refill ($n=13$).} Starting from a length-$2932$ snake of Echols, we refilled a window with a mixed-integer program constrained to close the path into a cycle, giving a $2916$-coil (Appendix~\ref{detail:sec:coils}).

\paragraph{Symmetry-restricted search (symmetric coils).} Every proven coil optimum in dimensions $2$--$8$ has a representative whose second half is the image of the first under an involutive cube automorphism. Fixing such a map halves the vertices to be chosen, so we adapted surgery (acting on vertex pairs) and Beam Anchor (building one half-path) to search within a fixed symmetry. These searches set the symmetric-coil records of $370$ in $Q_{10}$ and $734$ in $Q_{11}$; Appendix~\ref{app:coil-symmetry} gives both adaptations and their sweep across symmetry classes.

\section{Snakepit-in-the-Box}
\label{sec:snakepit}
Beyond their role in search, snakepits pose an extremal problem of their own. Let
\[
 \Lambda(n)=\max_{k\geq1}L(n,k)
\]
be the maximum edge count over all snakepits in $Q_n$. In dimensions $1$--$6$, snakepits cannot contain more edges than an optimal snake. Dimension $7$ is the first where they do: the maximum is $52$, versus $50$ for snakes, attained by four length-$13$ paths from the product construction in Section~\ref{sec:defs}. Appendix~\ref{app:snakepit-theory} gives the computer-assisted proof.

The same construction supplies a general lower bound: the product of two vertex-disjoint length-$13$ snakes in $Q_5$ has $2^{n-5}$ components and $13\cdot2^{n-5}$ edges in $Q_n$, so
\[
 \Lambda(n)\geq13\cdot2^{n-5}\qquad(n\geq5).
\]
In particular, this construction gives $104$ edges in $Q_8$ and $208$ in $Q_9$. A pit-mode variant of our annealer never exceeded these across $24$ two-hour runs in $Q_9$ (Appendix~\ref{detail:sec:fill}); their optimality is open.

Let $D(a)$ be the maximum of $e(A)+e(B)$ over vertex-disjoint snakepits $A,B$ in $Q_a$.
\begin{proposition}
\label{prop:density}
The edge density $\Lambda(n)/2^n$ converges, and its limit $\rho$ satisfies
\[
 \rho=\sup_a\frac{D(a)}{2^{a+1}},\qquad\frac{13}{32}\leq\rho\leq\frac12.
\]
\end{proposition}
\begin{proof}[Proof sketch]
Products of disjoint pairs make $D(a)/2^{a+1}$ nondecreasing in $a$, and Huang's sensitivity theorem \citep{huang2019} bounds every snakepit in $Q_n$, $n\geq5$, by $2^{n-1}$ vertices, so the ratio is at most $1/2$. Conversely, an optimal pit and its translate along a least-used coordinate can be made disjoint while losing at most $3\Lambda(n)/n$ edges, which squeezes $\Lambda(n)/2^n$ between $D(n-1)/2^n$ and $D(n)/2^{n+1}$ up to a factor $1-3/(2n)$. The two disjoint length-$13$ snakes in $Q_5$ give $13/32$; Appendix~\ref{app:snakepit-theory} gives the full proof.
\end{proof}
The deficit of a pair from density $1/2$ equals its unused vertices plus its components, divided by $2^{a+1}$. Improving on the $Q_5$ pair, or proving that no pair in any dimension does, is the benchmark's central asymptotic question. Table~\ref{tab:snakepit} lists the current values of $L(n,k)$.
\begin{table}[t]
\caption{The head of the Snakepit-in-the-Box table $L(n,k)$. Dashes are infeasible; $\Lambda(n)$ maximizes over all $k$. Unqualified entries are exact (Appendix~\ref{detail:sec:snakepit}); $\geq$ entries are achieved lower bounds.}
\vspace{0.2cm}

\label{tab:snakepit}
\centering\small
\setlength{\tabcolsep}{3.5pt}
\begin{tabular}{@{}crrrrrrrrr@{}}
\toprule
$n$ & $k=1$ & $2$ & $3$ & $4$ & $5$ & $6$ & $7$ & $8$ & $\Lambda(n)$\\
\midrule
$3$ & $4$ & $2$ & $0$ & $0$ & --- & --- & --- & --- & $4$ \\
$4$ & $7$ & $6$ & $5$ & $4$ & $2$ & $0$ & $0$ & $0$ & $7$ \\
$5$ & $13$ & $12$ & $12$ & $12$ & $10$ & $10$ & $8$ & $8$ & $13$ \\
$6$ & $26$ & $26$ & $25$ & $24$ & $24$ & $24$ & $24$ & $24$ & $26$ \\
$7$ & $50$ & $50$ & $50$ & $52$ & $51$ & $50$ & $49$ & $48$ & $52$ \\
$8$ & $98$ & $\geq 98$ & $\geq 100$ & $\geq 100$ & $\geq 100$ & $\geq 101$ & $\geq 102$ & $\geq 104$ & $\geq 104$ \\
$9$ & $\geq 191$ & $\geq 190$ & $\geq 191$ & $\geq 196$ & $\geq 197$ & $\geq 200$ & $\geq 200$ & $\geq 200$ & $\geq 208$ \\
\bottomrule
\end{tabular}
\end{table}

Entries should come with an explicit snakepit and the search budget; the full table, our examples and baseline code are available at \url{https://github.com/Math-AI-Caltech/Snake-in-the-Box}.

\section{Conclusion}
\label{sec:conclusion}
There are two key lessons from this work to highlight. First, the choice of state space and move is critical: cube surgery on snakepits reached $191$ from an empty cube in minutes on one CPU core, beating both single-vertex annealing on the same states and our best learned constructor with far more compute. Second, within construction, learning does help, and the form of supervision matters: a policy trained only on its own anchored comparisons reaches $190$ from the origin, where the same network trained by PPO stops at $184$ and the same search with a uniform policy stops at $178$.

Two concrete problems remain open: whether a length-$192$ snake exists in $Q_9$ (or the weaker question of whether a length-$191$ snake exists outside the known bank), and whether the product snakepits are optimal, that is, whether $\Lambda(8)=104$ and $\Lambda(9)=208$.
\clearpage
\section*{Ethics Statement}
This work studies an extremal combinatorial problem and involves no living subjects or personal data.  In particular, no snakes were harmed, placed in physical boxes, or operated on surgically. We do not anticipate specific societal harms arising directly from the methods or results presented here.

\section*{Reproducibility Statement}
All data and code supporting the reported results are publicly available at \url{https://github.com/Math-AI-Caltech/Snake-in-the-Box}. Appendix~\ref{app:beam-anchor} gives the Beam Anchor algorithm, its production settings and the settings of the reported runs; Appendix~\ref{app:sa} gives production surgery, annealing and lifting. Appendix~\ref{detail:sec:records} gives the record provenance, the MILP refill and the coil methods; Appendix~\ref{app:snakepit-theory} gives the snakepit proofs and benchmark calibration; and Appendix~\ref{app:data} prints a record snake in $Q_9$ and in $Q_{10}$. All record snakes, coils, snakepit examples and class banks are released there with verifiers.

\section*{The Use of Large Language Models}
We used large language models to help structure, draft, and revise the manuscript; identify and discuss related literature; and discuss methodological choices and the interpretation of results. We also used LLMs to help write and optimize code implementing the search methods and running experiments, with the aim of reducing experiment runtime and accelerating iteration. Computational witnesses are checked by explicit validity tests, as described in the paper. The authors double-checked LLM outputs by hand and take responsibility for the final manuscript, mathematical claims, code, experimental results, and other artifacts produced with LLM assistance.

\section*{Acknowledgements}
The project was sponsored by the Defense Advanced Research Projects Agency under cooperative agreement HR0011262E017, by the NSF AIMing grant 2522494, by the DRW Foundation, by a philanthropic gift from Les Kohn, and by a gift from Nebius Inc. M.T. is also supported by the U.S. Department of Energy (Grant No. DE-SC0011632) and by the Walter Burke Institute for Theoretical Physics and E.H. is supported by Schmidt Sciences. Additionally, this work was supported with Cloud TPUs from Google's TPU Research Cloud (TRC), with GPUs from the NVIDIA Academic Grant Program, by cloud computing resources provided by Nebius through the Research Program of Nebius Academy, and by Advanced Micro Devices, Inc. under the AMD University Program's AI $\&$ HPC Cluster. The content of the information does not necessarily reflect the position or the policy of the Government, and no official endorsement should be inferred.

\bibliography{references}
\bibliographystyle{iclr2027_conference}
\clearpage
\appendix
\section{Beam Anchor: Algorithm Details}
\label{app:beam-anchor}\label{detail:sec:beam-anchor}\label{app:beam-anchor-study}
This appendix gives the implementation details omitted from Section~\ref{sec:beam-anchor}, the training loop (Algorithm~\ref{detail:alg:beam-anchor}), the production settings (Table~\ref{tab:anchor-hparams}) and the settings of the reported runs.

\subsection{Anchor search}
\label{app:anchor-search}
\paragraph{Canonical prefixes.}
Searches and walks use the first-use coordinate ordering of Section~\ref{sec:defs}, so each prefix represents its class under coordinate permutations, moves that would only relabel are excluded from $A(p)$, and equivalent prefixes share one row of $\mathcal M$.

\paragraph{Quotas and scoring.}
An anchor receives one pool of $B=nb$ beams, $b$ being the nominal width per move; slot $i$ of the pool belongs to the $\lfloor im/B\rfloor$-th of the $m$ legal moves, so quotas differ by at most one. Each group starts from one live copy of its child with score zero. At every depth, a live state $s_i$ with score $\phi_i$ and legal action $u$ yields the candidate
\[
 \phi'_{i,u}=\phi_i+\log\pi_\theta(u\mid s_i),\qquad
 r_{i,u}=\phi'_{i,u}+T_{\mathrm{BA}}\,g_{i,u},\quad g_{i,u}\sim\mathrm{Gumbel}(0,1),
\]
with fresh noise at each depth. Each group keeps its own top candidates by $r$, up to its quota; survivors carry the clean score $\phi'$. States reached twice, identified by a hash of the visited set and head, keep the larger score. Searches run until every beam has completed; the capacity $H$ exceeds any snake in $Q_n$, so no search is truncated. If a group completes nothing beyond its child, the walk scores that move by $\ell(p)+1$ and writes no label. Because $\mathcal M$ stores maxima over searches made under successive policies, its entries are optimistic attained lengths, not unbiased values.

\subsection{Walks and temperatures}
\label{app:temperatures}
A cycle runs $P$ walks in lockstep, one batched search per step sharded across devices, until every walk has no legal move. During training, each walk starts from a prefix drawn uniformly from all canonical snakes of length $\ell_0$ (set to 10 in our experiments) to favor opening diversity; at search time with a frozen policy, walks start at the origin.

Three temperatures appear. Call an anchor \emph{shallow} if its length is at most $\ell_{\mathrm a}$ and \emph{deep} if it is at least $\ell_{\mathrm b}$, and let $f\in[0,1]$ be the fraction of training completed.
\begin{itemize}
\item The \emph{search noise} $T_{\mathrm{BA}}$ above is constant.
\item The \emph{walk temperature} $\tau_{\mathrm{walk}}$ in $\mu(a\mid p)\propto\exp(\widehat L(p,a)/\tau_{\mathrm{walk}})$ is $\tau_{\mathrm s}(f)$ at shallow anchors, $\tau_{\mathrm d}(f)$ at deep ones, and linear in the anchor length between $\ell_{\mathrm a}$ and $\ell_{\mathrm b}$, with
\begin{equation}
 \tau_{\mathrm s}(f)=3.5-1.5f,\qquad \tau_{\mathrm d}(f)=1.3-0.3f.
 \label{eq:walk-temperature}
\end{equation}
Shallow anchors run hot because the opening decides which region a walk explores and the outcomes barely separate openings there; the decrease in $f$ anneals exploration as the policy improves.
\item The \emph{teacher temperature} $\tau$ of Eq.~\ref{eq:distillation-loss} is $2$ at shallow anchors, $1$ at deep ones, linear between, and constant over training; it is independent of the walk temperature.
\end{itemize}

\subsection{Distillation}
Let $\mathcal S(p)\subseteq A(p)$ be the moves at $p$ with a recorded outcome. Training rows are prefixes with $|\mathcal S(p)|\geq2$; the target and the policy in Eq.~\ref{eq:distillation-loss} are both restricted to $\mathcal S(p)$, so an unmeasured legal move receives neither target mass nor gradient. The table holds at most $C$ rows under least-recently-used eviction. After each cycle's walks, the policy takes $U$ Adam steps on minibatches of $R$ rows with global gradient clipping; the first $W$ cycles collect only.

\begin{algorithm}[!ht]
\caption{Beam Anchor training (one cycle)}
\label{detail:alg:beam-anchor}
\begin{algorithmic}[1]
\Require policy $\pi_\theta$, table $\mathcal M$, walks $P$, width per move $b$, noise $T_{\mathrm{BA}}$, training fraction $f$, updates $U$, minibatch $R$
\State draw $P$ start prefixes (uniform over canonical length-$\ell_0$ snakes) and replay them into states
\While{some walk is live}
  \ForAll{live anchors $p_j$ in parallel}
    \State $\widehat{\boldsymbol L}_j\gets$ anchor search from $p_j$: one group per $a\in A(p_j)$, quotas summing to $nb$, run to completion
    \State $\mathcal M[p_j]\gets\max(\mathcal M[p_j],\widehat{\boldsymbol L}_j)$ elementwise \Comment{only the anchor row}
    \State sample $a_j\sim\mu(\cdot\mid p_j)$ at temperature $\tau_{\mathrm{walk}}(\ell(p_j),f)$;\quad $p_j\gets p_j\mathbin\Vert a_j$
  \EndFor
  \State freeze walks with no legal move
\EndWhile
\For{$U$ updates}
  \State sample $R$ rows: a populated depth uniformly, then a prefix $p$ with $|\mathcal S(p)|\geq2$ uniformly
  \State update $\theta$ by Eq.~\ref{eq:distillation-loss} with $\tau=\tau(\ell(p))$, restricted to $\mathcal S(p)$
\EndFor
\end{algorithmic}
\end{algorithm}

\subsection{Network details}
\label{app:policy-network}
Of the eight node features in Section~\ref{sec:beam-anchor}, the five indicator features are binary; the position along the snake is divided by the capacity $H$, and the counts of snake vertices at Hamming distance one and two are divided by $n$ and $\binom n2$. A linear layer with ReLU embeds the features into $\mathbb R^{64}$, and $K=6$ residual GIN layers with unshared weights and no normalization apply
\[
 h_v\leftarrow h_v+\mathrm{MLP}_k\Bigl((1+\epsilon_k)\,h_v+\sum_{u\sim v}h_u\Bigr),\qquad k=1,\ldots,K,
\]
with $u\sim v$ over the $n$ cube neighbors, $\epsilon_k$ a learned scalar initialized at zero, and $\mathrm{MLP}_k$ two ReLU layers of width $64$. The action head is a two-layer MLP of width $64$; illegal moves are masked to $-10^9$ before the softmax. 

\paragraph{Shrinking-ball inference.}
The action head reads the final embeddings of the head $h$ and its $n$ neighbors, and each layer propagates one hop, so with $B_r(h)$ the Hamming ball of radius $r$ around $h$ it suffices to compute $h^{(0)}_v$ on $B_{K+1}(h)$ and the output of layer $j$ on $B_{K+1-j}(h)$. The cube is vertex-transitive, so the index tables are static. For the six-layer $Q_9$ policy the balls have $502,466,382,256,130,46,10$ vertices: $1{,}792$ message-passing rows per state instead of $7\times512=3{,}584$, with identical logits up to summation order.
\paragraph{Critic head.}
The PPO baseline adds a value head to the same trunk, unused by Beam Anchor. With $\bar h_{\mathcal X}$ the mean of the final embeddings over a vertex set $\mathcal X$, and $\mathcal V$, $\mathcal V_{\mathrm{snake}}$, $\mathcal V_{\mathrm{avail}}$, $\mathcal V_{\mathrm{legal}}$ the sets of all, snake, available and legal next vertices,
\[
 V_\theta(p)=\mathrm{MLP}\bigl([h_{v},\,\bar h_{\mathcal V},\,\bar h_{\mathcal V_{\mathrm{snake}}},\,\bar h_{\mathcal V_{\mathrm{avail}}},\,\bar h_{\mathcal V_{\mathrm{legal}}},\,(\ell(p)+1)/H]\bigr),
\]
with $v$ the head, $\ell(p)$ the length of $p$, $H$ the capacity and a two-layer MLP of width $64$. The means are invariant under vertex relabelings, so $V_\theta$ is invariant under the automorphisms of $Q_n$. Since they need final embeddings of every vertex, the PPO baseline runs the full forward pass rather than the shrinking-ball one.

\begin{table}[!ht]
\centering\small
\caption{Beam Anchor production settings in $Q_9$. Length-dependent temperatures interpolate linearly between the shallow and deep thresholds; the walk temperature also decreases linearly over training (Eq.~\ref{eq:walk-temperature}).}
\label{tab:anchor-hparams}
\begin{tabular}{@{}lll@{}}
\toprule
& Parameter & Setting \\
\midrule
Search & width per move $b$; beams per anchor $B=nb$ & $192$; $1{,}728$ \\
& search noise $T_{\mathrm{BA}}$ & $0.25$, constant \\
& action-sequence capacity $H$ & $192$ (never binding) \\
Walks & walks per cycle $P$ & $16$ \\
& training start prefix length $\ell_0$ & $10$ \\
& shallow / deep thresholds $\ell_{\mathrm a}$, $\ell_{\mathrm b}$ & $10$, $20$ moves \\
& walk temperature, shallow / deep anchors & $3.5\to2.0$ / $1.3\to1.0$ over training \\
Distillation & teacher temperature, shallow / deep anchors & $2$ / $1$, constant \\
& table capacity $C$; eviction & $10^5$ rows; least recently used \\
& updates per cycle $U$; minibatch $R$ & $12$; $512$ \\
& warm-up cycles $W$; cycles & $5$; $300$ \\
& optimizer; learning rate; clipping & Adam ($\epsilon=10^{-5}$); $3\times10^{-4}$; $0.5$ \\
Network & GIN layers; width & $6$; $64$ \\
\bottomrule
\end{tabular}
\end{table}

\subsection{Settings of the reported runs}
\label{app:anchor-protocol}
All Beam Anchor runs use the settings of Table~\ref{tab:anchor-hparams}.
\begin{itemize}
\item \emph{Beam Anchor, trained policies.} Two policies were trained from random initializations with different seeds (Section~\ref{sec:beam-anchor}). Each was then frozen and run twice more, with different seeds, for $12$ hours on one eight-device MI250 node, with the walk temperature held at its end-of-training value ($f=1$ in Eq.~\ref{eq:walk-temperature}) throughout the run.
\item \emph{Beam Anchor, uniform policy.} The training configuration of Table~\ref{tab:anchor-hparams} with $\pi_\theta$ replaced by the uniform distribution over legal moves in every search and walk, run for $12$ hours on one MI210 device with a single seed. Best length plateaued at $178$.
\item \emph{PPO, same network.} The same trunk with its critic head, trained by proximal policy optimization \citep{ppo} on episodes that build one snake from the origin under the same legality mask and ordering, with reward one per move and no curriculum or imitation. Rollouts of $256$ environments for $128$ steps, $32$ minibatches, $4$ epochs per update, discount $0.999$, GAE $\lambda=0.95$, learning rate $6\times10^{-4}$ and entropy coefficient $0.1$, both annealed towards $4\times10^{-4}$ and $0.06$ over $10^8$ global steps. Four seeds, each run for $12$ hours on one MI210 device; the best snake is the longest rollout during training and plateaued at $184$ well before the end of training.
\end{itemize}

\begin{figure}[!ht]
\centering
\begin{tikzpicture}[x=0.5cm,y=1cm]
\draw[gray!60] (36.5,0) -- (54.5,0);
\foreach \x/\c in {38/2,39/8,42/12,43/25,44/44,45/8,46/1}{
  \filldraw[fill=gray!60,draw=black] (\x-0.4,0) rectangle (\x+0.4,{\c/44*1.3});
  \node[font=\tiny,above] at (\x,{\c/44*1.3}) {\c};
}
\foreach \x in {37,...,54}{\node[font=\tiny,below] at (\x,0) {\x};}
\node[font=\scriptsize] at (45.5,-0.55) {$\dmax$, uses of the most-used coordinate};
\end{tikzpicture}
\caption{Distribution of $\dmax$ over the $100$ length-$190$ classes found by Beam Anchor in $Q_9$. All lie in the continent's mode of the length-$190$ bank (Figure~\ref{fig:dmax}); none reaches the island's modes at $\dmax=52$ and $54$.}
\label{fig:ba-dmax}
\end{figure}
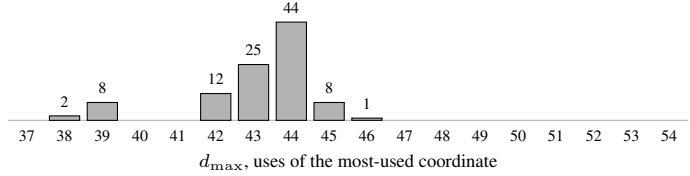

\section{Surgery, Annealing, and Lifting}
\label{app:sa}\label{detail:sec:modification}\label{detail:sec:construction}
\subsection{Production algorithms and settings}
\label{detail:sec:surgery}\label{detail:sec:sa}\label{detail:sec:othermoves}
The annealer's state is a snakepit, its energy is Eq.~\ref{eq:energy}, and its temperature follows a sawtooth with period $M$:
\begin{equation}
 T(t)=\max\left\{T_{\min},T_0\left(1-\frac{t\bmod M}{M}\right)\right\}.
 \label{detail:eq:temperature}
\end{equation}
The minimal move mixture (Algorithm~\ref{detail:alg:sa}) attempts surgery with probability $p_s$ and otherwise splits evenly between vertex addition and removal; production runs also use the endpoint moves of Section~\ref{sec:modification}.
\begin{algorithm}[!ht]
\caption{Cube surgery as one annealing proposal}
\label{detail:alg:surg}
\begin{algorithmic}[1]
\Require snakepit $F$ in $Q_n$, $n\geq4$, temperature $T>0$, penalty $\beta>0$
\State save $F_0\gets F$, including all metadata
\State draw $d\sim\mathrm{Unif}\{2,3,4\}$ and a random $d$-subcube $U$
\State delete every occupied vertex in $U$
\Repeat
  \State $r\gets0$
  \ForAll{unoccupied vertices $x$ of $U$, in a fresh random order}
    \If{$x$ has an occupied neighbor and adding $x$ preserves a snakepit}
      \State add $x$ to $F$ immediately; update metadata; $r\gets r+1$
    \EndIf
  \EndFor
  \ForAll{unoccupied vertices $x$ of $U$, in a fresh random order}
    \If{adding $x$ preserves a snakepit, including when $x$ has no occupied neighbors}
      \State add $x$ to $F$ immediately; update metadata; $r\gets r+1$
    \EndIf
  \EndFor
\Until{$r=0$}
\State $\Delta E\gets E(F)-E(F_0)$ using Eq.~\ref{eq:energy}
\If{$\Delta E>0$ and $\mathrm{Uniform}(0,1)\geq\exp(-\Delta E/T)$}
  \State restore $F\gets F_0$ exactly
\EndIf
\State \Return $F$
\end{algorithmic}
\end{algorithm}
\begin{algorithm}[H]
\caption{Minimal snakepit annealing chain}
\label{detail:alg:sa}
\begin{algorithmic}[1]
\Require dimension $n\geq4$, initial snakepit $F$, proposal budget $N$, $T_0,T_{\min},M,\beta,p_s$
\State $S_{\mathrm{best}}\gets F$ if $c(F)=1$, otherwise no snake yet
\For{$t=0,\ldots,N-1$}
  \State set $T$ by Eq.~\ref{detail:eq:temperature}
  \State draw $u\sim\mathrm{Uniform}(0,1)$
  \If{$u<p_s$}
    \State $F\gets$ Algorithm~\ref{detail:alg:surg}$(F,T,\beta)$
  \ElsIf{$u<p_s+(1-p_s)/2$}
    \State sample a cube vertex; if its addition is legal, test it using Eq.~\ref{eq:energy}
  \Else
    \State attempt to sample an occupied vertex; if found, test removal using Eq.~\ref{eq:energy}
  \EndIf
  \If{$c(F)=1$ and (no snake saved yet or $e(F)>e(S_{\mathrm{best}})$)}
    \State save a copy as $S_{\mathrm{best}}$
  \EndIf
\EndFor
\State \Return the saved snake, if any, after independent validation
\end{algorithmic}
\end{algorithm}

\begin{table}[!ht]
\centering\small
\caption{Nominal forest-annealer settings; run-specific overrides are recorded in the release manifests. The minimal mixture splits nonsurgery proposals equally between addition and removal; production endpoint-move probabilities depend on the recipe.}
\label{tab:production-settings}
\begin{tabular}{@{}ll@{}}
\toprule
Parameter & Nominal production setting\\
\midrule
Peak / minimum temperature & $2.0$ / $10^{-4}$\\
Fragmentation penalty & $\beta=3.2$, per-chain jitter $\pm0.30$\\
Surgery probability / dimension & $0.35$ / uniform $d\in\{2,3,4\}$\\
Reheating periods & $M\in\{64,128,256\}\times10^6$ moves\\
Chains / device; moves / launch & $131{,}072$; $2{,}000{,}000$\\
Stall threshold; bank reseeding & $30$ launches; probability $0.5$\\
Wiggle / deep regrowth & retract $1$--$6$ / $8$--$48$; grow up to $j+2$ / $K_{\rm removed}+12$\\
\bottomrule
\end{tabular}
\end{table}

Compound proposals (surgery, wiggle, deep regrowth) make one acceptance decision and restore the saved state exactly on rejection. Chains run in parallel on GPUs; a quarter of them run cool, and reseeding draws a random temperature phase. Chains save every single snake above a floor ($20,40,80,170,340,700,1400,2800$ in dimensions $6$--$13$), with deduplication. The first successful $Q_9$ chains started from the $200$-edge $Q_6$ product; the $208$-edge $Q_5$ product and banked snakes also led to $191$s.

\paragraph{Implementation.}\label{detail:sec:impl}
Occupancy is a bitset ($512$ bits in $Q_9$) with vertex, component and coordinate-use counts. Each endpoint stores its opposite endpoint, so a join that would close a cycle is detected in constant time. The same C core compiles for CPU and GPU and was fuzz-tested against brute force in dimensions $6$--$13$.

\subsection{Surgery comparison and annealing trace}
\label{app:surgery-comparison}
We compared two fixed annealing recipes in $Q_9$, each initialized from an empty snakepit and run for $3{,}600$ seconds on one dedicated CPU core for each of ten paired seed identifiers ($0$--$9$). Recipe A used single-vertex addition/removal with probabilities $0.5/0.5$; recipe B used surgery/addition/removal with probabilities $0.35/0.325/0.325$, sampling surgery dimension uniformly from $\{2,3,4\}$. Both used $\beta=3.2$ and the sawtooth temperature schedule with $T_0=2$, $T_{\min}=10^{-4}$ and period $64$ million attempted proposals. Both allowed fragmented states. Endpoint moves, auxiliary objectives, restarts, reseeding and random temperature phases were disabled; no witness bank was supplied.

Both recipes used the same binary, and the ten pairs ran concurrently on distinct cores of one CPU socket; all $20$ runs completed, with no reruns or tuning after the fact. Only a single snake counts toward the best length, and every reported snake was checked independently.

\paragraph{Results.}
The median final best snake length was $149$ edges (range $147$--$155$) for A and $188$ (range $186$--$191$) for B. Every paired difference favored B, with median $39$ edges and range $35$--$41$ (Table~\ref{tab:move-ablation}). Recipe B with seed $4$ found a validated $191$-edge snake after $158.484$ seconds and $302{,}413{,}283$ proposals.

\begin{table}[!ht]
\centering\small
\caption{All ten paired outcomes after $3{,}600$ seconds per recipe and seed in $Q_9$. A: add/remove; B: surgery/add/remove. Proposals count every attempt, including illegal and no-op proposals; full cycles count completed $64$-million-proposal periods.}
\label{tab:move-ablation}
\begin{tabular}{@{}rrrrrrrr@{}}
\toprule
& \multicolumn{3}{c}{Best snake edges} & \multicolumn{2}{c}{Proposals ($10^9$)} & \multicolumn{2}{c}{Full cycles} \\
Seed & A & B & B$-$A & A & B & A & B \\
\midrule
$0$ & $152$ & $188$ & $36$ & $16.015$ & $6.686$ & $250$ & $104$ \\
$1$ & $147$ & $186$ & $39$ & $15.300$ & $6.384$ & $239$ & $99$ \\
$2$ & $149$ & $187$ & $38$ & $16.151$ & $6.735$ & $252$ & $105$ \\
$3$ & $148$ & $188$ & $40$ & $16.194$ & $6.753$ & $253$ & $105$ \\
$4$ & $155$ & $191$ & $36$ & $16.145$ & $6.733$ & $252$ & $105$ \\
$5$ & $152$ & $187$ & $35$ & $16.142$ & $6.715$ & $252$ & $104$ \\
$6$ & $148$ & $187$ & $39$ & $16.148$ & $6.739$ & $252$ & $105$ \\
$7$ & $149$ & $189$ & $40$ & $16.131$ & $6.739$ & $252$ & $105$ \\
$8$ & $147$ & $188$ & $41$ & $16.202$ & $6.720$ & $253$ & $105$ \\
$9$ & $149$ & $188$ & $39$ & $16.161$ & $6.727$ & $252$ & $105$ \\
\bottomrule
\end{tabular}
\end{table}

\begin{table}[!ht]
\centering\small
\caption{Best-so-far validated single-snake edge lengths at the four prespecified checkpoints, across ten runs per recipe. Each checkpoint is recorded at the first completed proposal after its deadline.}
\label{tab:surgery-checkpoints}
\begin{tabular}{@{}rrr@{}}
\toprule
Elapsed checkpoint (s) & A median [range] & B median [range] \\
\midrule
$90$ & $143\ [143,147]$ & $181\ [175,184]$ \\
$240$ & $146\ [143,147]$ & $182.5\ [181,191]$ \\
$900$ & $147\ [144,155]$ & $185.5\ [184,191]$ \\
$3{,}600$ & $149\ [147,155]$ & $188\ [186,191]$ \\
\bottomrule
\end{tabular}
\end{table}

Table~\ref{tab:surgery-checkpoints} shows progress at the prespecified checkpoints. The equal-time comparison measures the performance of these two recipes: A completed $239$--$253$ temperature cycles, versus $99$--$105$ for B. Their proposal counts and temperature trajectories therefore differ; the comparison does not isolate surgery under a matched annealing trajectory or establish that surgery is necessary.

\paragraph{Annealing trace.}
Traced chains started from the $256$ isolated even-parity vertices show the intended cycle: hot phases carry more components and more edge changes, and cooling consolidates the pit. Few or no vertices stay occupied throughout a trace, and $120$ CPU-hours of such chains recovered $86$ known $190$ classes and $27$ known $191$ classes, from both families.

\subsection{Beam lifting}
\label{detail:sec:beam}

For a parent state $s$ of edge length $L$, let $R(s)$ contain the unvisited vertices whose visited neighbors, if any, consist only of the current head. These vertices have not yet been excluded by the path, although most are not legal next steps. Let $n_{\mathrm{legal}}(s)$ count the legal one-edge extensions. A child $s'$ receives score
\begin{equation}
 \operatorname{score}(s')=L+1+a_{\mathrm{reach}}\bigl(|R(s)|-n_{\mathrm{legal}}(s)\bigr)+T_{\mathrm{lift}}g(s'),
 \label{detail:eq:lift-score}
\end{equation}
where $g$ is a standard Gumbel draw and $T_{\mathrm{lift}}\geq0$ controls the selection noise. Equivalently, the deterministic part is the child's edge length plus $a_{\mathrm{reach}}|R(s')|$.

The count update is especially simple in a hypercube:
\[
 |R(s')|=|R(s)|-n_{\mathrm{legal}}(s).
\]
Once a new head is chosen, all previously legal neighbors of the old head leave $R$: the chosen vertex becomes occupied, while the others become forbidden by the old head. No two such neighbors are adjacent, so none can remain available through adjacency to the new head. Other vertices in $R$ remain available. The deterministic score is therefore shared by siblings and can be computed before materializing their states; noise distinguishes them.

\begin{algorithm}[!ht]
\caption{Lifting a primer from $Q_{n-1}$ into $Q_n$}
\label{detail:alg:beam}
\begin{algorithmic}[1]
\Require dimension $n$, primer snake, beam width $B$, noise scale $T_{\mathrm{lift}}$, $a_{\mathrm{reach}}$, seed
\State $\mathcal{S} \gets \{\text{the primer, placed in } Q_n\}$;\quad $\mathrm{best}\gets$ primer length
\While{$\mathcal{S}\neq\emptyset$}
  \State $\mathcal{C} \gets$ all one-edge extensions of states in $\mathcal{S}$ that remain induced paths
  \State score children by Eq.~\ref{detail:eq:lift-score}; remove duplicate states; $\mathcal{S} \gets$ the top $B$ children
  \State update $\mathrm{best}$ from generated children, if any
\EndWhile
\State \Return the longest snake found
\end{algorithmic}
\end{algorithm}

The record recipe uses widths $B=2^{16}$ over four seeds and $B=2^{18}$ over two, with $T_{\mathrm{lift}}=0.5$ and $a_{\mathrm{reach}}=1$. States are deduplicated by a $64$-bit hash of the visited set. Lifting all $131$ known $191$s recovered the same $12$ length-$379$ classes in $Q_{10}$; the tested $Q_8$ primers did not yield $190$ in $Q_9$.

\section{Records and Coils}
\label{detail:sec:records}\label{detail:sec:pipeline}
\subsection{Snake records}
The $Q_{12}$ route refined a $1470$ lift to $1474$, then re-lifted the improved $Q_{11}$ bank and annealed to $1476$. The $Q_{13}$ route refined $2916$ lifts to $2918$ and $2919$, then annealed a $2921$ seed from the genetic search below to $2924$. Class counts above $Q_9$ are harvest counts, not completeness results.

\paragraph{MILP window refill.}
Our genetic search refills a window of consecutive path vertices exactly. With the path outside the window fixed, let $W$ consist of the two cut endpoints $r$ and $s$ and the unused vertices adjacent to no fixed path vertex other than $r$ or $s$. With binary $x_v$ (vertex $v\in W$ chosen) and $y_{uv}$ (cube edge $uv$ inside $W$ used), and flows $f$ on directed edges, the refill solves
\begin{align*}
    \max&\sum_{v\in W}x_v\quad\text{s.t.}\quad
x_r=x_s=1,\quad
y_{uv}=x_ux_v\ (\text{linearized}),\\
&\sum_{u}y_{uv}\leq2x_v\ (\leq1\text{ at }r,s),\quad
\sum y=\sum x-1,
\end{align*}

with a single-commodity flow from $r$ that must reach every chosen vertex. Setting $y_{uv}=x_ux_v$ forbids chords; the degree, edge-count and connectivity constraints make the choice a single path from $r$ to $s$. Unlike the greedy refill of cube surgery, this solves the window exactly at the cost of one solver call. Scans sweep windows covering $1\%$, $5\%$, $10\%$ and $20\%$ of the path, ten per size, plus windows containing the endpoints. This produced the $2921$ seed in $Q_{13}$ and further record-length classes in $Q_{11}$.

\subsection{Coils}
\label{detail:sec:coils}
\paragraph{Closure.} A length-$L$ snake whose endpoints are at distance two closes into a length-$(L+2)$ coil when a common neighbor of the endpoints is unused and adjacent to no other path vertex. Of the $1{,}042$ length-$190$ classes in $Q_9$, $229$ have distance-two endpoints and $96$ close, all to the same $192$-coil.

\paragraph{Harvest and close.} No record-length snake in $Q_{10}$--$Q_{13}$ closes. We therefore ran the annealer warmer and with a lowered save threshold, and tested the harvested shorter snakes: $24$ million in $Q_{10}$, $19$ million in $Q_{11}$ and $14$ million in $Q_{12}$, giving the coil records $374$, $734$ and $1466$.

\paragraph{MILP closure.} The refill above adapts to coils by requiring the refilled window to join the two ends of the path into one cycle. Starting from a length-$2932$ snake of Echols whose endpoints were at distance $8$, it produced the $2916$-coil in $Q_{13}$.

\subsection{Symmetric coils}
\label{app:coil-symmetry}
\begin{definition}[Half-turn symmetry]
\label{def:sigma}
A coil $C=(v_0,\dots,v_{2h-1})$ in $Q_n$ is \emph{$\sigma$-symmetric} for an involutive cube automorphism $\sigma(x)=P(x)\oplus t$, where $P$ permutes coordinates, $P^2=I$ and $P(t)=t$, if $v_{i+h}=\sigma(v_i)$ for all $i$ (indices mod $2h$).
\end{definition}
Equivalently, the second half of the transition sequence is the first half with coordinates permuted by $P$; translation symmetry is the case $P=I$. For fixed $\sigma$, the half $v_0,\dots,v_{h-1}$ determines the coil: it must be an induced path disjoint from its image, joined to it by exactly the edges $v_{h-1}\sigma(v_0)$ and $\sigma(v_{h-1})v_0$. Every proven coil optimum in dimensions $2$--$8$ has a $\sigma$-symmetric representative.

\paragraph{Symmetric surgery.} The annealer keeps its occupied set invariant under $\sigma$: every addition or removal acts on the pair $\{v,\sigma(v)\}$, surgery clears $U\cup\sigma(U)$ and refills in pairs, and each paired edit gets one acceptance decision. Cycle components are allowed so that fragments can close up, and the energy is $\beta|c-1|-|V|$. Maps with $\min_x d(x,\sigma(x))<2$ are skipped. A state is saved only if it is a single induced cycle satisfying Definition~\ref{def:sigma}.

\paragraph{Half-path Beam Anchor.} Beam Anchor builds only the half-path: adding $u$ also occupies $\sigma(u)$, and a move to $u$ is legal only if $u$ and $\sigma(u)$ are unused and nonadjacent and the head is $u$'s only occupied neighbor. A move closes the coil when its only occupied neighbors are the head and $\sigma(v_0)$; the symmetric edge then closes the other side, giving a coil of length $2h$. A closure scores $10^5+2h$ and an open dead end scores $1$, so the anchor comparison favors moves that lead to long closures.

\paragraph{Sweep.} Symmetric surgery, run on every fixed-point-free class of $\sigma$ in $Q_8$--$Q_{11}$ with one worker per class, reached $96$, $192$, $372$ and $734$, matching the optimum or best known coil length in every dimension but $Q_{10}$. Half-path Beam Anchor reached $48$, $92$, $176$, $370$ and $730$ in $Q_7$--$Q_{11}$.

\subsection{Continent and island}
\label{app:landscape}
From a continent seed, add/remove, wiggle, surgery and deep regrowth recovered $96$, $125$, $127$ and $127$ of the $127$ continent classes, and never an island class. Only regrowth steered toward the island's coordinate profile crossed, after truncating a continent snake to length $130$. The same split appears at length $190$: $27\%$ of continent-type and $5\%$ of island-type $190$s extend to $191$, and no simple statistic we measured separates extendable $190$s from dead ends. The $190$ bank looked saturated at $1{,}041$ classes until a new scaffold-based sampler found a $1{,}042$nd, a dead end with $\dmax=49$. Generators trained on the bank (transformer, PatternBoost-style and GNN policies) likewise produced no $191$ outside it, and produced $191$s at all only when trained on extendable $190$s.

\clearpage
\section{Snakepit Benchmark and Mathematical Support}
\label{detail:sec:snakepit}\label{app:snakepit-theory}
\subsection{Bounds and small values}
Huang's sensitivity theorem shows that for $n\geq5$ any set of more than $2^{n-1}$ vertices induces a vertex of degree at least $\sqrt n>2$ \citep{huang2019}. Hence every snakepit has at most half the cube's vertices:
\begin{equation}
 L(n,k)+k\leq2^{n-1},
 \label{detail:eq:forest-bound}
\end{equation}
which also holds at $n=4$ by enumeration. With $k_{\min}(n)=\min\{k:L(n,k)=\Lambda(n)\}$, the exact initial terms are
\[
 \bigl(\Lambda(n)\bigr)_{n=1}^{7}=(1,2,4,7,13,26,52),
 \qquad
 \bigl(k_{\min}(n)\bigr)_{n=1}^{7}=(1,1,1,1,1,1,4).
\]
In $Q_7$ every optimum has exactly four components: $L(7,k)$ for $k=1,\ldots,8$ is $50,50,50,52,51,50,49,48$, and $48$ persists through $k=16$.

\subsection{Exactness in dimension 7}
\label{app:q7-proof}
The exact values for $n\leq6$ come from exhaustive enumeration. For $Q_7$ the proof is computer-assisted. For a split along coordinate $i$, write $L=L_0+L_1+c_i$, where each layer is a snakepit in $Q_6$ and $c_i$ counts cross edges. Suppose $L\geq53$ and choose $i$ least used, so $c_i\leq\lfloor L/7\rfloor$; the heavier layer then has $23\leq L_0\leq26$ edges. Each cross edge needs an endpoint or isolated vertex in that layer, so $k_0\geq\lceil c_i/2\rceil$ and $|S_0|=L_0+k_0\geq27$ vertices. A depth-first enumeration of all such layers containing a fixed degree-two pattern at the origin gives $4{,}576{,}964$ labeled sets and $5{,}041$ representatives under the automorphisms of $Q_6$. For each representative, a CP-SAT model \citep{ortools} enforcing induced degree at most two, acyclicity, the minimal split direction and the heavier-layer convention decides whether some second layer completes it to $53$ edges; all $5{,}041$ instances are infeasible, with no time-outs. Controls recover $\Lambda(6)=26$ through $Q_5$ layers and find feasible $Q_7$ completions at $52$.

At target $52$, the heavy layer has at least $26$ vertices and there are $12{,}199$ representatives. Completion with fixed component count excludes $k=2,3$ and $5$--$12$; the known snake optimum excludes $k=1$, and Eq.~\ref{detail:eq:forest-bound} excludes $k\geq13$. The $13$ feasible representatives at $k=4$ produce $169$ projected completions, reducing to $63$ classes under cube automorphisms. Every class consists of four length-$13$ paths with an edge-free coordinate; $62$ classes have two such coordinates and one has exactly one. The same procedure at targets $51$, $50$ and $49$, with $60{,}708$, $117{,}445$ and $496{,}524$ heavy-layer representatives, closes the remaining cells of the $Q_7$ row without time-outs.

\subsection{Proof of Proposition~\ref{prop:density}}
Let $d_n=D(n)/2^{n+1}$ and $u_n=\Lambda(n)/2^n$.

\emph{Products.} For vertex-disjoint $A,B$ in $Q_a$, the product of Definition~\ref{def:product} in $Q_{a+b}$ and the product with the roles of $A$ and $B$ swapped are disjoint, so $D(a+b)\geq2^bD(a)$ and $d_n$ is nondecreasing. A single product with $b=1$ gives $\Lambda(n)\geq D(n-1)$, that is, $u_n\geq d_{n-1}$. By Eq.~\ref{detail:eq:forest-bound}, $D(n)\leq2\Lambda(n)\leq2^n$, so $d_n\leq\frac12$.

\emph{Splitting.} Let $F$ be an optimal pit in $Q_n$ and choose a coordinate $i$ used by at most $\Lambda(n)/n$ of its edges. The translate $F\oplus e_i$ meets $F$ exactly in the endpoints of these $i$-edges. Deleting both endpoints of each one from the translate removes at most three edges per pair, since both have degree at most two and share the $i$-edge. The result $B$ is disjoint from $F$, so $D(n)\geq e(F)+e(B)\geq(2-3/n)\Lambda(n)$, that is, $u_n\leq d_n/(1-3/(2n))$.

Together, $d_{n-1}\leq u_n\leq d_n/(1-3/(2n))$ with $d_n$ nondecreasing and bounded, so $u_n$ and $d_n$ share the limit $\rho=\sup_a d_a\leq\frac12$. The disjoint length-$13$ snakes in $Q_5$ give $D(5)=26$ and $\rho\geq13/32$. \qed

For a disjoint pair with unused vertex set $R$, the deficit from $\frac12$ is exactly $(|R|+c(A)+c(B))/2^{a+1}$, since each component has one more vertex than edges. The $Q_5$ pair has deficit $6/64$. Its products attain $D(6)=52$ and $D(7)=104$, which are exact by $D(n)\leq2\Lambda(n)$; the first open base that could improve $13/32$ is $Q_8$, where a pair needs at least $209$ edges.

\subsection{Benchmark calibration}
\label{detail:sec:fill}
The pit-mode annealer records the best $e(F)$ for each component count it visits, with $\beta=1$ so that the energy is $-e(F)-1$. On $Q_8$, a penalty ramp from $1.2$ to $4.0$, surgery probability $0.38$ and wiggle probability $0.28$ reached $98$ in each of four $40$-minute runs from the empty pit, against $95$--$96$ with fixed $\beta=3.2$. Neither $32$ runs in $Q_8$ nor $24$ two-hour runs in $Q_9$ exceeded the product values $104$ and $208$.

\clearpage
\section{Record Snakes}
\label{app:data}
Each snake is written as its transition sequence: starting from the origin, the $j$th entry is the coordinate flipped at step $j$, with coordinates numbered from $0$. A length-$191$ snake in $Q_9$ from the continent, with DCV $(39,38,35,29,26,9,9,4,2)$:
\begin{quote}\footnotesize\ttfamily\raggedright
0 1 2 0 3 1 4 5 0 6 2 1 0 6 5 0 2 6 3 1 4 3 2 1 0 3 5 0 2 3 6 2 1 0 6 2 3 6 0 7 4 0 6 3 2 6 0 1 2 6 3 2 0 6 4 8 3 0 2 6 3 1 5 3 2 6 3 1 2 0 3 6 2 3 4 0 2 3 6 2 0 1 6 2 3 6 0 7 5 0 6 3 2 6 0 1 2 6 3 2 0 6 5 2 3 6 2 0 1 3 4 1 2 0 1 3 2 6 1 0 2 1 5 6 2 1 0 2 6 3 0 2 1 0 6 7 4 6 0 1 2 0 6 3 2 0 1 2 6 0 4 8 1 6 2 0 1 3 5 1 2 0 1 3 2 6 1 0 2 1 4 6 2 1 0 2 6 3 0 2 1 0 6 7 5 6 0 1 2 0 6
\end{quote}
A length-$379$ snake in $Q_{10}$, with DCV $(81,63,62,61,60,19,16,12,4,1)$:
\begin{quote}\footnotesize\ttfamily\raggedright
0 1 2 3 0 2 4 1 0 2 3 0 1 5 3 0 2 3 6 4 2 3 0 2 1 6 0 2 3 0 7 5 1 2 4 5 3 4 2 0 3 2 1 4 3 2 0 5 8 7 1 0 2 4 1 2 3 0 1 2 4 1 7 5 4 1 2 4 3 0 2 3 5 4 2 3 0 2 1 8 0 1 4 2 3 6 4 2 1 4 7 0 1 4 2 1 0 3 2 1 4 2 0 1 5 0 3 7 4 3 2 1 4 2 0 3 4 2 1 4 3 5 1 4 2 1 8 0 2 1 4 2 0 3 4 2 1 7 0 1 2 3 0 2 4 1 0 2 3 7 6 5 4 3 0 7 1 2 4 1 0 3 4 1 2 4 6 3 2 4 1 2 3 0 7 4 3 0 2 3 4 1 2 3 0 2 4 3 7 1 9 7 4 1 2 4 3 0 2 4 1 2 3 4 7 3 0 5 1 0 2 4 1 2 3 0 1 2 4 7 6 4 1 7 0 1 2 3 0 2 4 1 0 2 3 0 1 5 3 0 2 3 6 4 2 3 0 2 1 6 0 2 3 0 7 4 3 2 1 4 2 0 3 2 1 4 5 6 3 0 4 1 2 0 3 4 6 3 2 4 1 2 0 8 1 2 4 1 0 2 3 5 4 3 2 1 4 2 0 3 4 2 1 7 0 4 5 0 1 4 2 1 0 3 2 1 4 2 0 1 5 2 1 4 2 0 3 4 6 3 2 4 1 2 0 3 1 2 4 7 3 4 2 0 3 2 1 4 3 2 0 5 1 3 7 1 0 3 2 0 1 4 2 0 3 2 1 0 7 2 0 3 2 1 4 3 6 4 2
\end{quote}

\end{document}